\documentclass{aa} 
\usepackage{graphics,latexsym,amssymb,times,psfig}
\begin{document}
\title{Extremely hard GRB spectra prune down the forest of emission models}
\author{Giancarlo Ghirlanda \inst{1} \and Annalisa Celotti \inst{1} \and 
Gabriele Ghisellini \inst{2}}
\offprints{G. Ghirlanda; ghirland@sissa.it}  
\institute{SISSA/ISAS, via Beirut 2-4, I--34014 Trieste, Italy;
\and Osservatorio Astronomico di Brera, via Bianchi 46, I--23807 Merate, Italy.}
%
%\date{Received ....; accepted ....}
\date{}

\titlerunning{Extremely hard GRBs}
\authorrunning{G. Ghirlanda, A. Celotti \& G. Ghisellini}

\abstract{We consider the evidence for very hard low energy spectra
  during the prompt phase of Gamma-Ray Bursts (GRB).  In particular we
  examine the spectral evolution of GRB 980306 together with the
  detailed analysis of some other bursts already presented in the
  literature (GRB 911118, GRB 910807, GRB 910927 and GRB 970111), and
  check for the significance of their hardness  (i.e. extremely
  steep spectral slopes below the $EF_{E}$ peak) by applying
  different tests.  These bursts, detected by the Burst And
  Transient Source Experiment (BATSE) in the $\sim 30$ keV -- 2 MeV
  energy range, are sufficiently bright to allow time resolved
  spectral studies on time intervals of the order of tenths of a
  second.  We discuss the hard spectra of these bursts and their
  evolution in the context of several non--thermal emission models,
  which all appear inadequate to account for these cases. The
  extremely hard spectra,  which are detected in the early part of
  the BATSE light curve, are also compared with a black body
  spectral model: the resulting fits are remarkably good, except for
  an excess at high energies (in several cases) which could be simply
  accounted for by the presence of a supra--thermal component.  The
  findings on the possible thermal character of the evolving spectrum
  and the implications on the GRB physical scenario are considered in
  the frameworks of photospheric models for a fireball which is
  becoming optically thin, and of Compton drag models, in which the
  fireball boosts ``ambient" seed photons by its own bulk motion. Both
  models, according to simple estimates, appear to be
  qualitatively and quantitatively consistent with the found spectral
  characteristics, although their possible caveats are
  discussed.
 }

\maketitle

%________________________________________________________________

\section{Introduction}

Since the discovery of these mysterious and puzzling cosmic
explosions, the time resolved spectral analysis of Gamma-Ray Bursts
(GRBs) has been a testing ground--floor for the models proposed for
the $\gamma$--ray prompt emission (Ford et al. \cite{Ford a}, Crider
et al. \cite{Crider aph}).  In general GRBs have a continuous
curvature spectrum which is usually characterized (Band et
al. \cite{Band}) by two smoothly joining components defining a peak
(in a $EF_E$ representation of the power per unit logarithmic energy
interval) at energies around 200 keV.  The low energy asymptotic
spectrum can be modeled as a power law function in the photon
distribution $N(E) \propto E^{\alpha}$ (i.e.  $E F_E \propto E^{\alpha
+2}$), where E indicates the photon energy. The high energy spectral
component can be described in most cases by a steep power law
$E^{\beta}$ (with $\beta < \alpha$) or an exponential cutoff.  The
typical values of $\alpha$ and $\beta$ are widely distributed around
--1 and --2.5 respectively (Preece et al.  \cite{Preece2000}; see also
Fig.~1).

In particular, different authors (Crider et al.  \cite{Crider a},
Preece et al.  \cite{Preece1998a}) have stressed the importance of
studying the low energy component of GRB spectra which is one of the
better determined and accessible observables.  Crider et
al. (\cite{Crider}) showed that the low energy spectrum evolves in
time and that the distribution of the spectral index $\alpha$ extends
above -1 and in some cases can be $\alpha \ge 1$.

The most popular radiative process proposed for interpreting GRB
spectra is synchrotron emission by relativistic electrons in intense
magnetic fields (Rees \& M\'esz\'aros \cite{Rees}; Katz \cite{Katz};
Tavani \cite{Tavani}). Optically--thin synchrotron theory
predicts (Katz \cite{Katz}) that the low energy photon spectrum can
not be harder than $N(E)\propto E^{-2/3}$. Nonetheless, different
authors (Crider et al. \cite{Crider}, Preece et
al. \cite{Preece1998a}, Preece et al. \cite{Preece2002}) have found
violations of this limit in many observed spectra. Following these
inconsistencies between the standard synchrotron model and
observations, alternative scenarios have been proposed, such as 
jitter radiation (Medvedev \cite{Medvedev}); synchrotron emission from
particles with an anisotropic pitch angle distribution (Lloyd \&
Petrosian \cite{Lloyd}, \cite{Lloyd a}); thin/thick synchrotron
emission from a stratified region (Granot, Piran \& Sari
\cite{Granot}); synchrotron self--Compton or inverse Compton off
photospheric photons (M\'esz\'aros \& Rees \cite{Meszaros}); Compton
drag (Lazzati et al. \cite{Lazzati 2000}; Ghisellini et al.
\cite{Ghisellini}); Comptonization of low energy photons by thermal
or quasi--thermal particles (Liang et al. \cite{Liang}; Ghisellini \&
Celotti \cite{Ghisellini a}).  The physical parameters of these models
can be tuned and combined to justify the principal observed temporal
and spectral characteristics of GRBs, producing a considerable number
of plausible spectral shapes.  However, the most extreme (in
this case hardest) low energy spectra and their evolution can be
used to rule out or constrain some of these possibilities.

Another relevant aspect of GRB spectra is their possible thermal
character. In fact, the fireball model (Goodman \cite{Goodman},
Paczynski \cite{Paczynski}) naturally predicts thermal radiation when
the fireball becomes transparent. The lack of observational evidence of
thermal spectra motivated the proposal of the internal shock model
(Rees \& M\'esz\'aros \cite{Rees}) in which the fireball energy could
be efficiently extracted and transformed into radiation with a
non--thermal spectrum as observed in most bursts.  Nonetheless,
evidence for possible thermal spectra in some GRBs has been reported
by Preece (\cite{Preece2001}) for GRB 970111 and Schaefer et al. (poster
P-56 presented at the Rome 1998 Whorkshop on GRBs, private
communication). Furthermore Blinnikov et al. (\cite{Blinnikov})
propose that also the spectra observed in most GRBs could be
interpreted as superposition of black body spectra with different
temperatures. From a theoretical point of view thermal emission from
the fireball is expected if the dominant acceleration agent is
internal pressure (Daigne \& Mochovitch \cite{Daigne}
2002). Alternatively, magnetic acceleration (e.g. by Poynting flux)
would determine a non--thermal spectrum (Drenkhahn \& Spruit
\cite{Drenkhahn}, but see also Ruffini et al. \cite{Ruffini}). The
analysis of spectra with a possible thermal character can therefore
have important implications for the understanding of the nature of the
fireball acceleration mechanism.

To the aim of testing the validity/generality of the proposed models
we have then searched among the 156 GRBs of the published spectral
catalog of Preece et al.  (\cite{Preece2000}) the bursts with a low
energy time resolved spectrum harder than $N(E)\propto E^{0}$, which
represents a limit for most of the above models (see Sec. 2). We found
two extremely hard bursts: GRB 911118 (BATSE trigger 1085),
whose peak energy evolution has been reported in  the spectral
catalog by Ford et al. (\cite{Ford a}), and a new case, GRB 980306
(trigger 6630).  These two GRBs are carefully studied here in
terms of their low energy spectral hardness. Their spectral 
evolution on timescales of few tenths of a second is presented
in Sec. 4.1, 4.2, particularly regarding the slope of the low
energy spectral component.  We also found GRB 910807 (trigger
647) and GRB 970111 (trigger 5773) whose spectral properties have
already been presented in the literature by Crider et
al. (\cite{Crider}, \cite{Crider aph}, \cite{Crider b}) and Frontera
et al. (\cite{Frontera}) (only for GRB 970111). We included also GRB
910927 (trigger 829) reported by Crider et al. (\cite{Crider},
\cite{Crider b}) which is not present in the spectral catalog of
Preece et al. (\cite{Preece2000}). These bursts have been re-analyzed
in order to test the reliability of their low energy spectral hardness
(Sec. 4.4). In fact, we present the tests that we performed in order
to determine the statistical robustness of the spectral hardness, and
in particular a model independent approach consisting in the
comparison of each spectrum with a template one of given hardness
(Sec. 4.3).  The evidence that, at least at the beginning of the
bursts, the spectra are extremely hard suggested their comparison
with a black body model. This part of the analysis is presented in
Sec. 4.5.  The comparison with the low energy spectral limits
predicted by different models (briefly recalled in Sec. 2 together
with the previous evidence of hard spectra) is the content of the
discussion (Sec. 5), where also tentative interpretations of the
initial quasi--thermal spectral evolution, in the context of the hot
fireball model and of the the Compton drag model, are presented.
We draw our conclusions in Sec. 6.

%____________one column table___________________________________
\begin{table}    
\caption[]{Low energy limiting photon spectral index $\alpha$
[i.e. $N(E) \propto E^\alpha$] for various emission models.  For
clarity the indices for the energy spectrum $F(E)$ and for its
$E F_{E}$ representation are also reported.  }
\label{alfa_models}
\begin{tabular}{|cccl|}
\hline 
% \noalign{\smallskip}
$\alpha$  & $\alpha+1$ & $\alpha+2$ & \\
% \hline 
% \noalign{\smallskip}
$N(E)$ & $F(E)$ & $EF_{E}$ & model/spectrum \\
% $$\mathrm{\frac{phot}{cm^{2}\           s\          keV}}$$          &
% $$\mathrm{\frac{keV}{cm^{2}\  s}}$$  & $$\mathrm{keV(\frac{keV}{cm^{2}
% s})}$$  & model  \\  
% \noalign{\smallskip} 
\hline  
% \noalign{\smallskip}
-3/2 & -1/2 & 1/2 &  Synchrotron emission with cooling \\  
-1   &  0   &  1  & Quasi--saturated Comptonization \\
-2/3 & 1/3  & 4/3 & Instantaneous synchrotron \\ 
0    & 1    & 2   & Small pitch\ angle/jitter\\
     &      &     & inverse Compton by\ single $e^{-}$  \\ 
1    & 2    & 3   & Black Body \\  
2    &  3   & 4   & Wien  \\
% \noalign{\smallskip} 
\hline 
\end{tabular}
\end{table}

%_______________________________________________________________

%__________________________________________________________________

\section{Spectral Slopes and Emission Models} 

\begin{figure}
\resizebox{\hsize}{!}{\includegraphics[110pt,368pt][540pt,698pt]{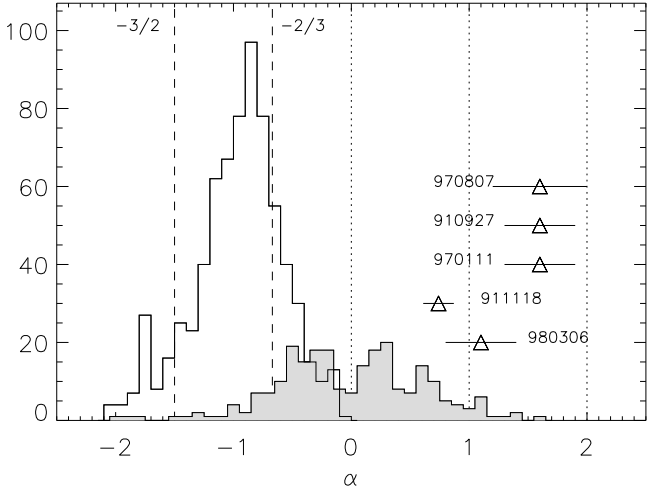}}
\caption[]{Low energy spectral index distribution from Ghirlanda,
Celotti \& Ghisellini (\cite{Ghirlanda}). The vertical lines represent
the limits of the emission models reported in Tab. \ref{alfa_models}
and plotted in Fig. \ref{fig1}.  The \textit{triangles} represent the
maximum values found from the spectral analysis of the bursts
presented in this work along with their 90\% confidence interval
(horizontal bars).  The gray histogram shows the distribution of the
low energy spectral index found from the time resolved spectral
analysis of the bursts presented in the text.}
\label{fig0}
\end{figure}

\begin{figure}
\resizebox{\hsize}{!}{\includegraphics[91pt, 334pt][540pt,700pt]{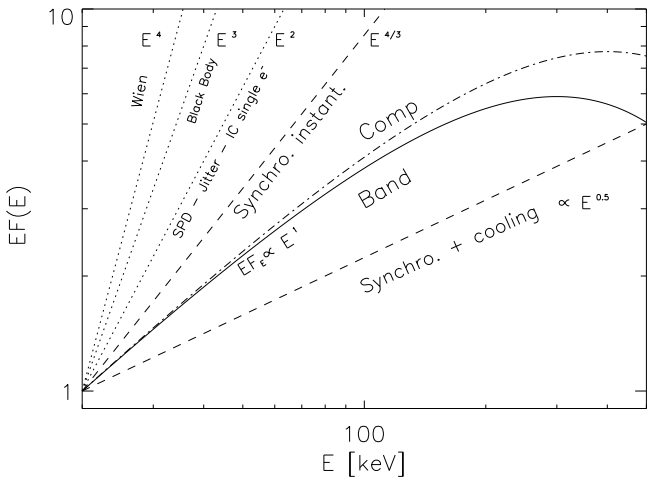}}
\caption[]{Low energy spectral limits for different emission models
(see Tab.\ref{alfa_models}).  Examples of the BAND (\textit{solid
line}) and COMP model (\textit{dot--dashed line}) with the low energy
photon spectral index fixed at --1 [i.e.  $EF_{E}\propto E^{1}$] 
are also represented for comparison. }
\label{fig1}
\end{figure}

The distribution of the low energy photon spectral index $\alpha$
obtained from the time resolved spectral analysis of a sample of
bright BATSE bursts (Ghirlanda, Celotti \& Ghisellini
\cite{Ghirlanda}) is reported in Fig.~\ref{fig0}.  The majority of GRB
spectra (Preece et al.  \cite{Preece2000}, Ghirlanda et
al. \cite{Ghirlanda}) have a low energy power law spectral index $-3/2
\le \alpha \le -2/3$, i.e.  within the limits predicted by the
optically thin synchrotron model (Fig.~\ref{fig1}, \textit{dashed
lines}; see also Katz \cite{Katz}), but there is a non negligible
fraction of bursts ($\sim$15\%, see also Preece et al.
\cite{Preece2000}) with a low energy spectrum harder than
$N(E)\propto E^{-2/3}$.  The spectral analysis of a sample of GRBs by
Crider et al.  (\cite{Crider}) also revealed that in 60\% of
their cases the spectrum evolves with time. Moreover, the
spectrum integrated over the pulse rise phase is harder than
$E^{-2/3}$ in 40\% of their bursts and can be as hard as
$N(E)\propto E^{1}$ (i.e.  $EF_E \propto E^{3}$).  For comparison the
distribution of the $\alpha$ parameter for the hard spectra of
the bursts considered in this work is also reported (gray
histogram) in Fig.  \ref{fig0}. Note that they contribute to
extend the distribution toward positive $\alpha$ values up to $\sim
1.5$.

%___One column figure = The alfa distribution _______________________
%____________________________________________________________________

Let us here just recall that this evidence is hardly reconcilable
with the simplest formulation of the synchrotron model and some
alternatives have been proposed to account for these observations
within the frame of this emission process (Papathanassiou
\cite{Papathanassiou}, Granot, Piran \& Sari \cite{Granot}).  Lloyd \&
Petrosian (\cite{Lloyd}) propose a scenario in which electrons have a
small pitch angle distribution (SPD in Fig.  \ref{fig1}), extending
the range of possible low energy spectral indices produced via
synchrotron to the limit $\alpha\sim 0$.  The same limiting slope can
be obtained in the ``jitter" radiation theory, which is based on
synchrotron emission in a non uniform magnetic field with
inhomogeneities on length scales smaller than the electron gyro-radius
(Medvedev \cite{Medvedev}).  Slopes even harder (i.e. $\alpha\sim 1$)
may instead correspond to thermal radiation, such as a portion of a
black body spectrum (M\'esz\'aros \& Rees \cite{Meszaros}) or
saturated Comptonization spectra with Wien peaks which can be as hard
as $\alpha\sim 2$ (Liang et al.  \cite{Liang}, Ghisellini \& Celotti
\cite{Ghisellini a}, Ghisellini et al.  \cite{Ghisellini}).\\
A detailed discussion of these models and their comparison with our
observational findings are presented in sec.~6. As reference for the
following sections we schematically summarize the model predictions in
Tab.  \ref{alfa_models}, and display in Fig. \ref{fig1} their typical
low energy limiting slopes in  a $EF_{E}$ representation.

%___One column figure = The spectral limits  _______________________
%____________________________________________________________________

\section{Data Analysis} 

We have analyzed the Large Area Detector (LAD) High Energy Resolution
Burst (HERB) data which have a high count rate, due to the LAD large
effective area, and are suited for the spectral analysis of GRBs for
which a moderate energy resolution is sufficient to study the typical
broad band continuum (Preece et al. \cite{Preece1998}).

We selected the data from the most illuminated detector which has the
highest signal to noise ratio (S/N), and maintained the
instrumental time binning corresponding (in the best cases) to single
time resolved spectra accumulated for 128 ms, to follow the spectral
variations on the smallest possible timescale (thanks to the high
S/N there was no need to rebin the data before fitting).  Each
spectrum has been fitted from $E_{\rm min}\sim 28$ keV to $E_{\rm
max}\sim 1800$ keV which define the typical energy window for the LAD
data (Preece et al.  \cite{Preece1998}, \cite{Preece2000}).

The spectral analysis has been carried out using SOAR (\cite{Ford
1993}) as a quick look tool for the spectral evolution and then
XSPEC {\it(v11.1)} for the individual spectral fitting of time resolved
spectra.The background, to be subtracted to each time
resolved spectrum, has been calculated as the average over a selected
number of background spectra accumulated before and after the trigger.

In the data analysis we have used the standard forward folding
  technique to fit the model spectral functions (convolved with the
  detector response matrix) to the observed spectra. As explained in
  Sec.4, we also used the instrumental spectra in the comparison with
  a simulated spectrum in order to verify the robustness of our
  results.

\subsection{Fitting models} 

We fitted the most commonly adopted GRB spectral models: the BAND
(Band et al.  \cite{Band}) and COMP model, represented by a smoothly
connected double power law or by a single power law ending with an
exponential cutoff, and the sharply connected double power law model
(BPLW).  The choice of these models is motivated by the fact that they
characterize the low energy part of the spectrum with a spectral
parameter (i.e. the spectral index $\alpha$) which can be simply
compared with the predictions of the different {emission
processes}.  For a detailed description of these spectral functions
see Preece et al.  (\cite{Preece2000}).

Due to the extreme hardness of the low energy spectrum of these
bursts, we decided also to verify if thermal, black body like,
emission was consistent with the low energy data.  We considered only
the time resolved spectra which have a low energy spectral index
$\alpha \ge 0.5$ when fitted with the BAND or COMP models.  This
indicative value, which is softer than $\alpha=1$ (i.e. the
Rayleigh-Jeans limit of a black body spectrum) has been chosen to: a)
account for typical errors in the determination of $\alpha$ with the
BAND or COMP models; b) include the possibility that the spectrum
could be softer than $\alpha=1$, if its peak is at low energies, due
to the spectral curvature (see also Sec.4.5).

The result of each fit was then considered acceptable if the reduced
$\chi^{2}_{\rm r}$ was lower than 1.5 for typically $\sim$ 110
degrees of freedom (dof): in fact the statistical probability of
having a better fit is around 0.5 if the reduced $\chi^{2}_{\rm r}$ is
around 1, but a limiting value of $\chi^2_{\rm r}\sim 1.5$ is
suggested if one considers that $\chi^{2}_{\rm r}$ depends on the
quality of the data (Bevington \& Robinson 1992).  In addition we
visually inspected the data--to--model ratio to look for possible
systematics (i.e., sequence of points significantly above or below
unity).  We discarded the black body as a good fit when an excess of
flux occurred at low energies (i.e.  in the $E^2$ part of the
spectrum), while we allowed for (a moderate) deviation from unity at
high energies (i.e. in the exponential part).  This latter choice
reflects the possible (or even likely) situation of having a
``supra--thermal" tail to a Maxwellian distribution in quasi--thermal
plasmas, or a multi--temperature black body emission as proposed
by Blinnikov et al. (\cite{Blinnikov}).

\section{Results} 

Sec. 4.1 deals with the low energy spectrum of GRB 980306.  In
particular we focus on the early phase of this burst when the spectrum
is extremely hard and report the time evolution of its low energy
spectral component.  The detailed spectral analysis of GRB 911118
(also reported by Ford et al. \cite{Ford a} only in terms
of the time evolution of the peak energy) is presented in Sec. 4.2
with emphasis (again) on the low energy spectral index evolution.  In
Sec.  4.3 we describe the tests we performed on the raw data to check
for the significance of these results on the spectral hardness.
In Sec. 4.4 we briefly discuss other 3 bursts (GRB 910807, GRB 910927
and GRB 970111) with hard spectra at low energies which were
already presented in the literature. We have re-analyzed them,
applying the tests described in Sec.4.3, to verify their low energy
hardness. In particular for GRB 910807 we have extended the time
analysis beyond the first 5 s that were previously studied by
Ryde \& Svensson (1999). The results relative to the black body fits
are reported separately in Sec. 4.5.

\subsection{\bf GRB 980306}

\begin{figure}
\resizebox{\hsize}{16cm}{\includegraphics{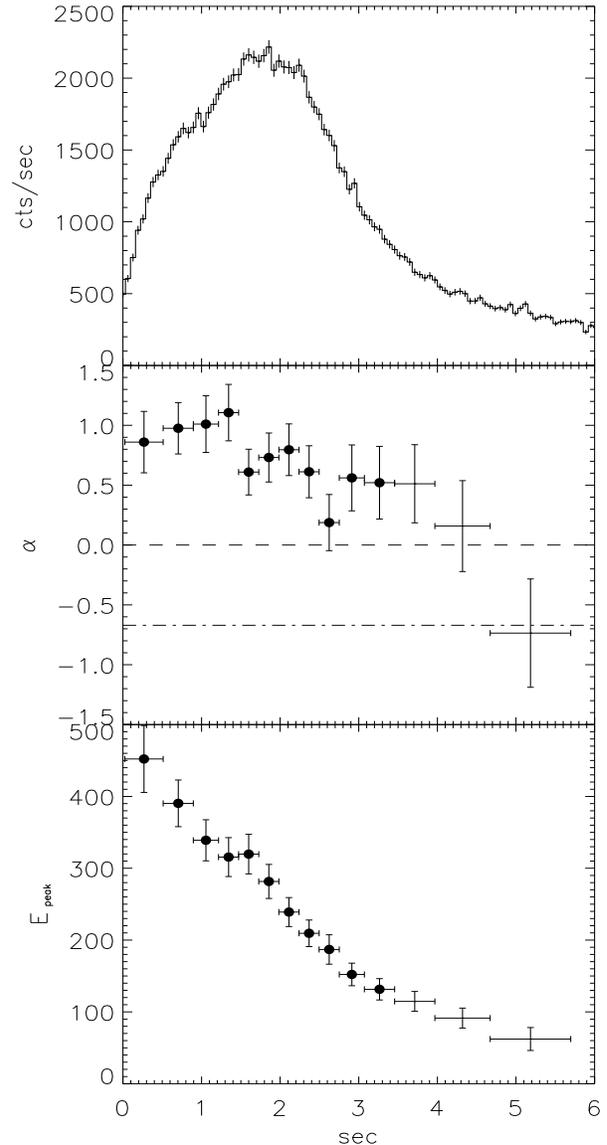}}
\caption[]{ Light curve and best fit parameters for GRB 980306. Top
panel: light curve on the 64 ms timescale, integrated in the range 110
-- 320 keV and summed over all triggered detectors.  Mid panel:
low energy power law spectral index ($\alpha$).  Horizontal lines mark
the limits $\alpha=-2/3$ ({\it dot-dashed}) and $\alpha=0$ ({\it
dashed}).  Bottom panel: peak energy $E_{\rm peak}$ of the $EF_{E}$
spectrum.  Error bars represent the 90\% confidence intervals on
the best fit parameters.  The filled circles indicate the spectra
which have been fitted also with a black body model (see
Tab. \ref{Black_body} and Sec. 4.5). }
\label{fig7}
\end{figure}

GRB 980306 represents a new case of hard burst.  It is a single peak
burst (Fig. \ref{fig7}, \textit{top panel}) with duration of $\sim 6$
s, peak flux of $(17.2\pm 0.4)$ phot cm$^{-2}$ s$^{-1}$ at
$t_{peak}\sim 2.2$ s and a total (100--300) keV fluence of $\sim
10^{-5}$ erg
cm$^{-2}$.\footnote{http://cossc.gsfc.nasa.gov/batse/4Bcatalog/index.html}
This burst was already present in the Preece et
al. (\cite{Preece2000}) spectral catalogue but, as it had been fitted
with a BPLW model only, it resulted softer than what we have
found by fitting its spectra with the COMP model. In fact, as already
noticed in Ghirlanda et al. (\cite{Ghirlanda}), a fit of a model with
a sharp break, like the BPLW, to a smoothly curved spectrum, although
can still give acceptable fits, tends to underestimate the low energy
spectral hardness.
 
We present, for the first time, its hardness evolution, during the
first 6 s, divided in 14 time resolved spectra.  In particular, we
focus on the extremely hard low energy spectral component. The
spectral fits with the different models show that the spectra are
better represented by the COMP model, since the high energy spectrum
is very steep (i.e. steeper than a power law with spectral index
$\beta=10$, the limit of the fitting function). The time evolution of
the low energy spectral index $\alpha$ and of the peak energy
$E_{peak}$ are reported in Fig.\ref{fig7} (middle and bottom panels,
respectively).

%___Onecolumn figure = The burst light curve_______________________
%____________________________________________________________________

The hardness evolution of a burst is typically described by how the
spectral index $\alpha$ and $E_{\rm peak}=(\alpha +2)E_0$ change in
time (here $E_{\rm peak}$ is the peak energy of the $EF_{E}$ spectrum
and $E_0$ is the characteristic energy of the exponential cutoff).
The evolution of $\alpha$ (mid panel of Fig. \ref{fig7}) indicates
that, in the rising phase of the pulse, the maximum hardness
($\alpha=1.1\pm 0.2$) is reached during the interval $t \le 1.5$
s. The first 4 spectra are in fact consistent with a non evolving
spectrum (the low energy spectral index can be considered equal to its
average value $\langle \alpha \rangle=0.73 \pm 0.03$ at 92\%
confidence level for the first 1.5 seconds).  After this time the
spectrum softens, but it remains harder than $\alpha=0$.  We stress
that the low energy spectrum is significantly (more than 3$\sigma$)
harder than $\alpha=0$ for each of the first 7 time resolved spectra
(between $t=0$ and $t=2.24$ s). For this specific burst the evolution
is hard--to--soft, as shown by the peak energy (bottom panel of Fig.
\ref{fig7}) decreasing in time from $\sim$450 keV to $\sim$50 keV,
consistently with the $E_{peak}$ evolution reported by Ryde \&
Svensson (\cite{Ryde2002}).

In view of the discussion on the spectral models (sec. 6) we also
checked the inadequacy of the optically thin synchrotron model by
fitting the BAND spectrum with the low energy power law slope fixed at
$-2/3$. As expected, the residuals at low energies show a systematic
sign and the reduced $\chi^{2}_{\rm r}$ is larger than 2
indicating that this model is not a good representation of the data.

\subsection{\bf GRB 911118}

\begin{figure}
\resizebox{\hsize}{16cm}{\includegraphics{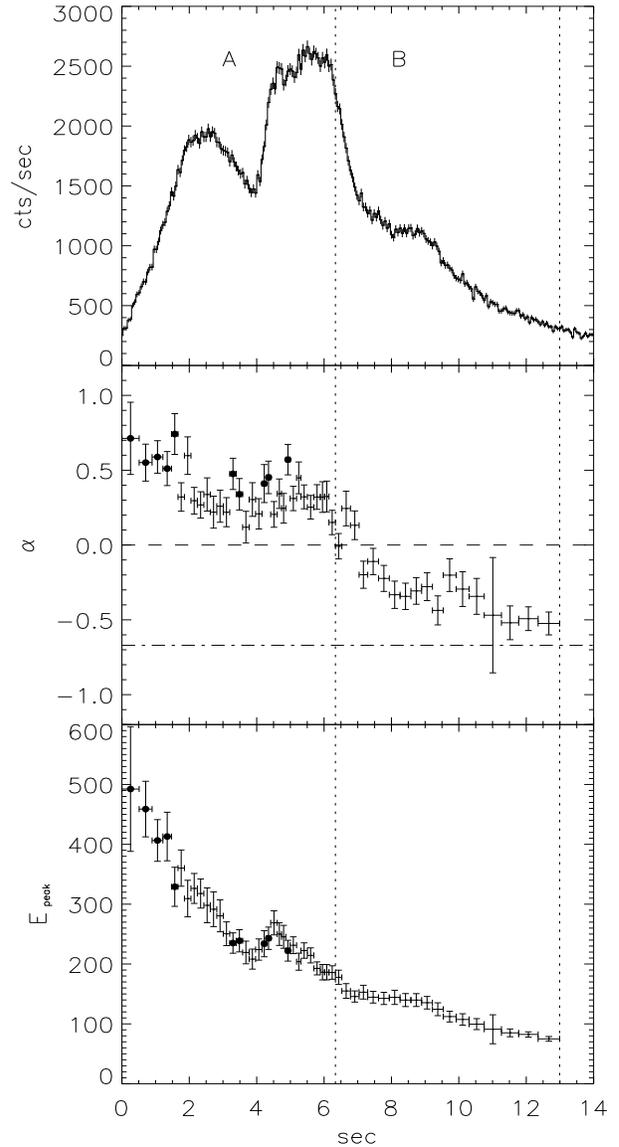}}
\caption[]{ The light curve and time evolution of the best fit
parameters of the BAND model for GRB 911118.  Top panel: light curve
on the 64 ms timescale in the energy range 110 -- 320 keV, summed over
all the triggered detectors. The vertical \textit{dotted} lines
represent the boundaries of the phases in which the spectrum is harder
than $E^{0}$ (phase A) and $E^{-2/3}$ (phase B).  Mid panel: low
energy power law spectral index $\alpha$, with the limits at
$\alpha=-2/3$ ({\it dot--dashed}) and $\alpha=0$ ({\it dashed}).
Bottom panel: peak energy of the $EF_{E}$ spectrum. The filled circles
represent the spectra which have been fitted also with a blackbody
model (see Tab. \ref{Black_body} and Sec. 4.5). }
\label{evolution_grb}
\end{figure}

The light curve of GRB 911118, reported in Fig.~\ref{evolution_grb}
(\textit{top panel}), has two main peaks which partially overlap.  It
is a typical long burst with total duration $T_{90}=19.2 \pm 0.1$ s
and a background subtracted peak flux of ($30.6\pm0.8$) phot cm$^{-2}$
s$^{-1}$ at $t_{\rm peak}=6.08$ s.  Its fluence, in the 100--300 keV
energy range, is $(27.8\pm 0.1)\times 10^{-6}$ erg cm$^{-2}$.$^{1}$

The spectral analysis that we present covers the first 13 s
after the trigger time (for a total of 50 spectra), in which the time
resolved spectrum evolves dramatically with an excursion of $\Delta
\alpha \sim 1$ and $\Delta E_{peak} \sim 400$ keV. The best fit is
obtained with the BAND model.  The low energy spectral index is
$\alpha >0$ for the first $\sim 6$ s (32 spectra, phase A of
Fig.~\ref{evolution_grb}, {\it mid panel}) and decreases with time
(the spectrum softens), but remains harder than --2/3 for the time
interval up to the end of phase B (see Fig. 4).  After $t\sim$13
s, the spectrum has a typical $\alpha < -2/3$ slope.  The hardest
spectrum (at $t\sim 1.5$ s after the trigger) has $\alpha=0.74\pm
0.13$. The statistical significance of a positive value of
$\alpha$ for the spectra of phase A is high, being $>$ 3$\sigma$ for
most of the individual spectra (23/32).  The peak energy evolution
reported in Fig. \ref{evolution_grb} ({\it bottom panel}) is
consistent with what reported by Ford et al.  (\cite{Ford a}).
According to these fits, the 18 spectra belonging to phase B have
$-2/3 \le \alpha \le 0$ and their evolution is of the
``hard--to--soft" kind, i.e.  the peak energy decreases in time
(becoming smaller than 200 keV in phase B).  Fig.~\ref{evolution_grb}
shows a possible correlation between the peak energy $E_{peak}$
and $\alpha$. It has been pointed out (Lloyd \& Petrosian
\cite{Lloyd}) that such correlations could result from the effect of
the curvature of the fitted model when the peak energy is particularly
low (e.g. below $\sim 100$ keV and close to the low energy spectral
threshold). However in the case of GRB 911118 we can exclude this
effect because the peak energy is between 500 and 200 keV during
phase A (Fig.\ref{evolution_grb}). 

%___One column figure = The burst spectral evolution _______________________
%____________________________________________________________________

\begin{figure}
\resizebox{\hsize}{!}{\includegraphics{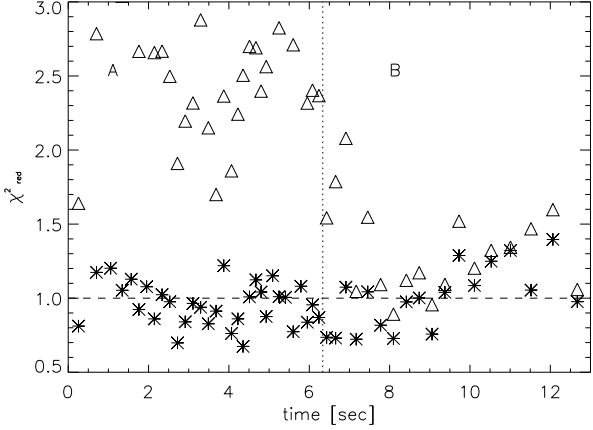}}
\caption[]{ Reduced $\chi^{2}_{\rm r}$ (for typical 110 dof) of the
  fits with different models to the spectra of GRB 911118: BAND model
  (\textit{asterisks}), BAND model with the low energy spectral index
  fixed at --2/3 (\textit{triangles}).  }
\label{fig5}
\end{figure}

Again, we also tried to fit the simplest optically thin synchrotron
model (by fixing the low energy spectral index of the BAND model to
--2/3).  In Fig.~\ref{fig5} we report, for comparison, the values of
the reduced $\chi^{2}_{r}$ for all spectra.  The best fit is given by
the BAND model (\textit{asterisks}) while the ``synchrotron'' 
case ({\it triangles}) is inadequate to fit these spectra 
because its $\chi^{2}_{\rm r}$ is typically $\ge 2$ (for $\sim 110$
degrees of freedom) during phase A.  The value of $\chi^{2}_{r}$ for
the synchrotron model becomes marginally acceptable ($<1.5$) for 
most spectra in phase B.

%___One column figure = The chi^2 _______________________
%____________________________________________________________________

We conclude that the time resolved spectra of this burst are best
represented by the BAND model with the 4 fit parameters free to vary
and that most of the spectra of phase A have $\alpha>0$ at a $\ge
3\sigma$ level.

\subsection{Tests on the Spectral Results}

Considering the relevance of the extreme hardness of these bursts for
the comparison with the models, we performed different checks to
verify if and at what confidence level they can be considered harder
than the predictions of the proposed spectral models (as in
Tab. \ref{alfa_models}).  In the following we report on the tests
performed on the above GRB spectra and their results.

\begin{figure}
\resizebox{\hsize}{!}{\includegraphics{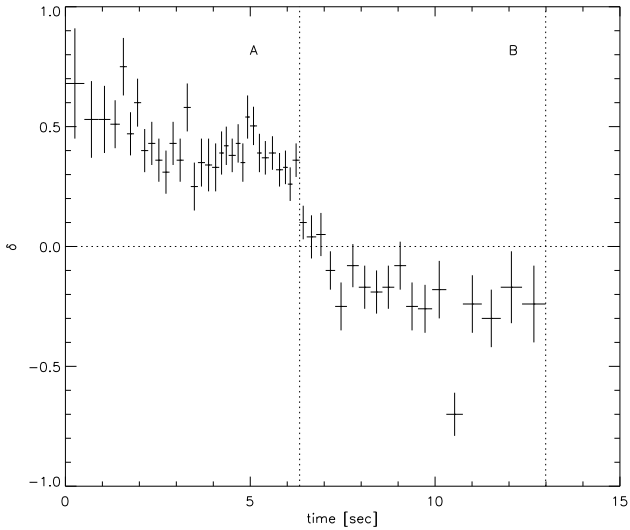}}
\caption[]{GRB 911118 - Slope of the power law fitted to the ratio
between the observed time resolved spectra and the template spectra
(simulated with $\alpha=0$ and $E_{break}=E_{0}(fit))$. Labels A and B
refer to the two phases of GRB 911118 (Fig.\ref{evolution_grb}) and
the horizontal dotted line is the slope of the reference spectrum.}
\label{fig8}
\end{figure}

\begin{figure}
\resizebox{\hsize}{!}{\includegraphics{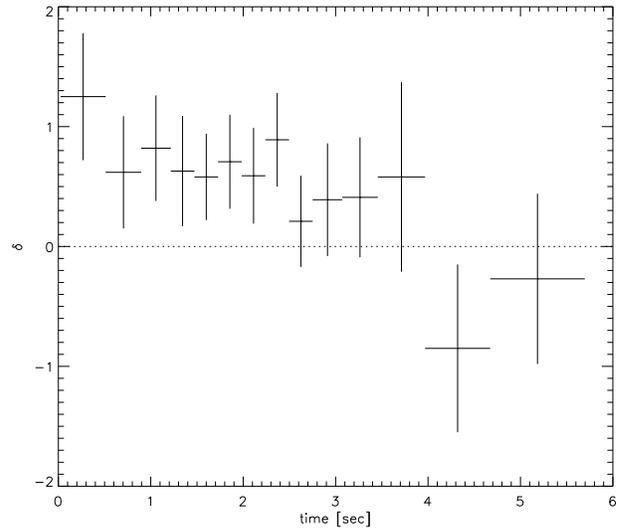}}
\caption[]{As Fig. \ref{fig8} for GRB 980306.}

%itted power law  slope of the  ratio between
%the observed time resolved  spectra and the template spectra simulated
%with $\alpha=0$ at low energies.}
\label{fig9}
\end{figure}

\begin{itemize}
\item {\it reduced fitting energy range}: we eliminated from the
fitting energy range the high energy channels to test if they
substantially influence the low energy fitted spectral slope.  We
restricted the analysis to the range 28--600 keV (the upper limit
being the maximum peak energy found among the bursts presented) and
compared the results with those obtained by fitting the spectra on the
entire energy window (i.e. up to 1800 keV).  We found that the best
fit spectral parameters are consistent within their errors and the
values of $\alpha$ inferred from the limited energy range case are
still inconsistent with $\alpha=0$.  Moreover, Preece et
al. (\cite{Preece1996}) found a soft low energy component in some
bursts through the analysis of other data type extending below the
standard LAD threshold of 28 keV.  This component could make the
fitted low energy spectral slope softer than it really is and the fit
results for the $\alpha$ parameter presented here should be considered
as lower limits thus reinforcing the evidence of extremely hard
spectra. We tested this possibility by fitting the spectra with a low
energy threshold at 50 and 100 keV. The comparison with the fits on
the entire energy range (28 - 1800 keV) showed consistent spectral
parameters within their 90\% confidence intervals.
  
\item {\it background spectrum}: if the background spectrum were
characterized by any feature or absorption edge at low energies
(e.g. dependent on some electronic failure or anomalous observational
conditions) this might determine particularly hard low energy power
laws in the background subtracted spectra.  In order to exclude this
possibility we computed several backgrounds selecting different time
intervals, within the same GRB, and found that the background spectrum
does not depend on the time interval selected for its estimate.
Moreover, we also fitted the spectra adopting different backgrounds
calculated for other bursts: the results indicate that $\alpha$ is not
affected by anomalous features in the background spectrum.

\item \textit{detector response}: the detector response matrix, if
incorrectly calculated, could determine a wrong set of best fit
parameters.  We believe this is unlikely as, by fitting the spectra
with the detector response matrix (DRM) associated with another burst
but for the same LAD, the results are consistent within their 90\%
confidence level. Moreover, although we used the data of the mostly
illuminated detector, we also tried to fit the data from the other
three detectors that were triggered by these bursts. Although the
lower signal to noise ratio in these ``secondary'' detectors resulted
in a lower statistical significance of the best fit parameters, these
fits confirmed the extremely hard low energy spectral slope of these
bursts at least in the second best illuminated detector. For the
remaining two detectors the results are consistent, although the low
statistics does not allow to confirm the hardness. Note also that
Crider et al. (1997) showed that combining different detectors can
make the low energy spectral component softer than it really is.
  
\item \textit{comparison with a simulated spectrum}: the most direct
and robust way to verify if the deconvolved spectra are really as hard
as we found is to compare them with a reference spectrum whose shape
is well defined. For every time resolved spectrum that we analyzed, we
thus simulated a reference (template) spectrum with a low energy power
law slope $\alpha \sim 0$ and a break energy fixed at the value
obtained from the fit of the observed spectrum.  We divided channel by
channel the observed spectrum by the template and then analyzed the
ratio (as long as the observed spectrum is harder than the template
one their ratio at low energies increases with energy).  In order to
quantify the hardness we then fitted such ratio {$R(E)$, where
$E$ represents the energy,} with a power-law model $R(E)=A\
E^{\delta} + B$. The fitted index $\delta$ is reported in Fig.
\ref{fig8} and \ref{fig9} for GRB 911118 and GRB 980306, respectively.
We note that during phase A (the first 6 s after the trigger) of GRB
911118 each of the time resolved spectra is at least 3$\sigma$ harder
than a flat spectrum (i.e. with the same level of confidence that was
found with the direct fitting method).  In the case of GRB 980306 this
test gives a fitted ratio with a poor level of confidence ($<2\sigma$)
for each spectrum (with only two spectra at $>2\sigma$) because of the
lower S/N due to count statistics and to the propagation of errors in
dividing the observed spectrum by the template one. This
significance is increased if the first 8 spectra are rebinned in time
(although the time resolved information is lost) which gives a
value of $\delta>0$ at 3.7$\sigma$. Furthermore (see also below)
also in the case of GRB 980306 the significance of $\delta>0$
(Fig. \ref{fig9}) is higher when considering the sequence of $\delta$
values as a group (since $\delta$ remains systematically positive).

\end{itemize}

As a final consideration let us stress the fact that in the case
of GRB 911118 (Fig. \ref{evolution_grb}) and GRB 980306
(Fig. \ref{fig7}) there is a minority of hard spectra (9 and 4,
respectively) which have $\alpha>0$ at a level of significance lower
than 3$\sigma$.  Nonetheless the fact that a considerable number of
subsequent spectra (32 for GRB 911118 and 12 for GRB 980306) in the
first few seconds have a spectrum harder (at more than $3\sigma$) than
$E^{0}$ increases the level of significance of the result, if they
are considered as a whole.

%%___One column figure = The burst light curve _______________________
%____________________________________________________________________
%%___One column figure = The burst light curve _______________________
%
%____________________________________________________________________

\subsection{Other hard bursts}

There are other bursts with extremely hard low energy spectra
for a major part of their evolution, which have already been reported
in the literature.  In particular, we found three bursts with very
hard spectra if fitted by the BAND function. In order to have
homogeneous results and verify their robustness we have then
re--analyzed their spectra and applied the tests described in
Sec.4.3. Let us summarize their properties.

\begin{itemize}
  
\item GRB 910807 has a peak flux of $(7.2\pm 0.5)$
phot cm$^{-2}$ s$^{-1}$.  The spectrum is harder than $\alpha=0$ for
the first $\sim$5 s, and remains quite constant ($\sim 0$) during the
overall burst, which lasts about 28 s, as already pointed out by Ryde
\& Svensson (\cite{Ryde}).  The peak energy instead shows a
tracking pattern correlated with the flux and hardens in
correspondence of the second peak (at $t=12$ s) as already
reported by Borgonovo \& Ryde (\cite{Borgonovo}).

%%___One column figure = 910807 evolution _______________________
\begin{figure}
\resizebox{\hsize}{16cm}{\includegraphics{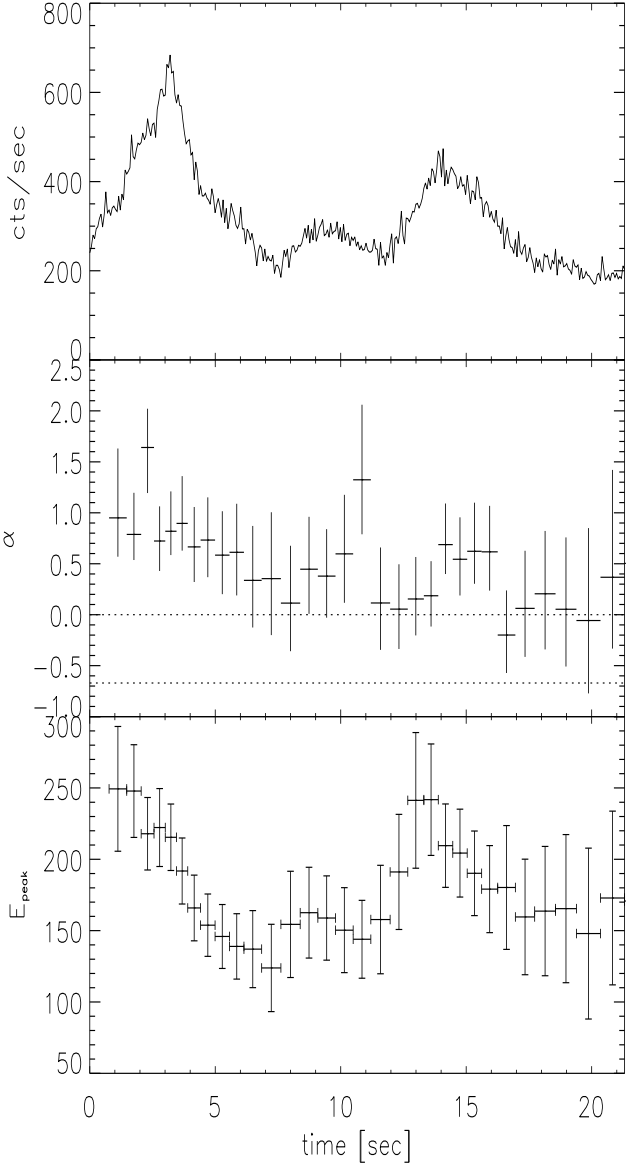}}
\caption[]{GRB 910807 spectral evolution. Same as Fig. \ref{fig7}}
\label{fig10}
\end{figure}
%____________________________________________________________________

The complete spectral evolution is reported in Fig. \ref{fig10}.  The
hardest spectrum has $\alpha=1.6\pm 0.5$ at 90\% confidence level.
This value is somewhat harder, but consistent with what quoted by
Crider et al.  (\cite{Crider b}). A (successful) check for the
hardness of the first few spectra of this burst has been performed
using the last method described in the previous section.

\item GRB 910927 (Crider et al.  \cite{Crider b}, \cite{Crider}) and
  GRB 970111 (Crider et al.  \cite{Crider aph}; Frontera et al.
  \cite{Frontera}) are other two cases of extremely hard spectra with
  a low energy power law spectral index as high as $\alpha= 1.6\pm
  0.3$.  GRB 970111 was studied (Frontera et al. \cite{Frontera}) also
  including the WFC--$Beppo$SAX data.  Its extremely hard low energy
  spectrum was confirmed to extend down to 5 keV. Unfortunately, the
  lack of spectral resolution of the GRBM instrument above 40 keV does
  not allow to study the complete spectrum, especially at its peak.

\end{itemize}

In addition we found in the spectral catalogue of Preece et al.
(\cite{Preece2000}) other 3 bursts (BATSE triggers 1974, 2855 and
6350) with $\alpha>0$, but at a low significance level ($<1\sigma$)
and/or for a short interval of the light curve.  Note that also these
cases contribute to the $\alpha >0$ tail of the spectral index
distribution reported in Fig. \ref{fig0}.

\subsection{Black body spectra}

As anticipated we performed fits on the hardest spectra also with a
black body function to test if the emission is consistent with a
thermal model.  In fact, the first spectra of all the bursts discussed
in this work are typically harder than $\alpha = 0.5$ and thus could
be consistent with the Rayleigh--Jeans part of a black body. Note that
the fit could however be unacceptable in terms of $\chi^2_{\rm r}$ if
the curvature around the peak (i.e. typically in the range 100 - 300
keV) were broader than a black body one.

%_______________________ TABELLONA DEI BB _____________________________
\begin{table*}[h]
\caption{Black body fits.} \label{Black_body}
{\scriptsize \begin{center}
\begin{tabular}{ccccccccc}
\hline\hline
GRB  &   $t_{\rm start}$  & $t_{\rm stop}$  & kT  & N & $\chi^2_{\rm r}$(dof) & F & $\alpha$ & $\chi^{2}_{\rm r,\alpha}$(dof)\\
     &   sec              &    sec          & keV & erg/cm$^2$ sec  & &     phot/cm$^2$ sec &   &   \\
\hline
%      &  &  &  &  &  & & & \\  
980306 & 0.024  & 0.512 & $104^{+4.8}_{-4.8}$ & $46^{+2.7}_{-2.8}$ & 1.15(107) & 8.4 & $0.86\pm 0.15$ & 1.09(109) \\
       & 0.512  & 0.896 & $ 93^{+4.6}_{-3.4}$ & $63^{+3.2}_{-2.8}$ & 1.27(107) & 12.9 & $0.98\pm 0.13$ & 1.21(104) \\
       & 0.896  & 1.216 & $81^{+2.9}_{-2.8}$ & $63^{+2.8}_{-2.8}$ & 1.27(107) & 14.7 & $1.01\pm0.13$ & 1.2(106) \\                          
       & 1.216  & 1.472 & $78^{+2.7}_{-2.6}$ & $68^{+3.1}_{-3.0}$ & 1.19(103) & 16.6 & $1.11\pm 0.14$ & 1.19(102) \\
       & 1.472  & 1.728 & $72^{+2.5}_{-2.5}$ & $70^{+3.1}_{-3.0}$ & 1.44(105) & 18.4 & $0.61\pm 0.12$ & 1.12(104) \\
       & 1.728  & 1.984 & $65^{+2.1}_{-2.1}$ & $66^{+2.7}_{-2.6}$ & 0.86(107) & 19.10 & $0.73\pm 0.12$ & 0.7(106) \\
       & 1.984  & 2.24  & $57^{+1.6}_{-1.8}$ & $59^{+1.7}_{-2.6}$ & 1.33(104) & 19.09 & $0.8\pm 0.13$ & 1.16(103) \\
       & 2.24   & 2.496 & $49^{+1.5}_{-1.4}$ & $52^{+1.8}_{-1.8}$ & 1.26(104) & 19.3 & $0.61\pm 0.13$ & 0.96(103) \\
       & 2.496  & 2.752 & $  42^{+1.3}_{-1.4}$ & $39^{+1.3}_{-1.4}$ & 1.39(101) & 16.6 & $0.19\pm 0.14$ & 0.8(100) \\
       & 2.752  & 3.072 & $ 37^{+1.1}_{-1.1}$ & $30^{+0.9}_{-0.9}$ & 1.45(104) & 14.1  & $0.56\pm 0.17$ & 1.2(103) \\
       & 3.072  & 3.456 & $ 33^{+1.0}_{-1.0}$ & $23^{+0.7}_{-0.8}$ & 1.23(107) & 12.14 & $0.52\pm 0.19$ & 1.0(106)\\
      &  &  &  &  &  & & & \\       
911118 & 0.000  & 0.512 & $114^{+8.0}_{-8.0}$ & $23^{+2.0}_{-3.0}$ & 0.95(111) & 3.92 & $0.71\pm 0.24$ & 0.8(109) \\
       & 0.512  & 0.896 & $98^{+4.0}_{-5.0}$ & $39^{+2.0}_{-3.0}$ & 1.39(109) & 7.61 & $0.55\pm 0.12$ & 1.1(107) \\
       & 0.896  & 1.216 & $87^{+3.0}_{-5.0}$ & $52^{+3.0}_{-3.0}$ & 1.51(105) & 11.28 & $0.59\pm 0.11$ & 1.2(103) \\
       & 1.216  & 1.472 & $89^{+3.0}_{-4.0}$ & $69^{+4.0}_{-4.0}$ & 1.45(107) & 14.78 & $0.51\pm 0.11$ & 1.0(105)\\
       & 1.472  & 1.664 & $76^{+3.0}_{-3.0}$ & $68^{+4.0}_{-3.0}$ & 1.35(105) & 16.93 & $0.74\pm 0.13$ & 1.1(103)\\
       & 3.2    & 3.392 & $52^{+1.0}_{-2.0}$ & $65^{+2.0}_{-2.0}$ & 1.52(104) & 22.76 & $0.5\pm 0.1$ & 0.9(102)\\
       & 3.392  & 3.584 & $52^{+1.3}_{-2.0}$ & $59^{+2.0}_{-2.0}$ & 1.51(107) & 20.61 & $0.57\pm 0.1$ & 0.87(105)\\
      &  &  &  &  &  & & & \\ 
910807 & 0.768  & 1.472 & $71^{+4.6}_{-4.6}$ & $9.4^{+0.8}_{-0.8}$ & 1.10(113) & 2.51 & $0.95\pm 0.29$ & 1.2(111)\\
       & 1.472  & 2.048 & $61^{+3.3}_{-3.5}$ & $11^{+0.4}_{-0.9}$ & 1.29(113) & 3.3  & $0.78\pm 0.21$ & 1.0(111)\\
       & 2.048  & 2.56  & $59^{+2.8}_{-2.6}$ & $17^{+0.8}_{-0.8}$ & 1.05(112) & 5.32 & $1.64\pm 0.25$ & 0.8(110) \\
       & 2.56   & 3.008 & $57^{+2.2}_{-2.3}$ & $19^{+1.2}_{-2.3}$ & 0.78(111) & 6.12 & $0.72\pm 0.19$ & 0.84(109) \\
       & 3.008  & 3.456 & $53^{+2.2}_{-2.1}$ & $21^{+0.9}_{-0.9}$ & 0.94(109) & 7.36 & $0.82\pm 0.18$ & 0.95(107) \\
       & 3.456  & 3.904 & $53^{+1.8}_{-1.9}$ & $24^{+1.0}_{-1.0}$ & 1.00(112) & 8.43 &$0.89\pm 0.21$ & 1.06(110) \\
       & 3.904  & 4.416 & $47^{+1.8}_{-1.7}$ & $19^{+0.9}_{-0.8}$ & 1.09(111) & 7.42 &$0.67\pm 0.23$ & 0.97(109) \\
       & 4.416  & 4.992 & $41^{+1.6}_{-1.7}$ & $14^{+0.6}_{-0.6}$ & 1.06(113) & 6.05 &$0.58\pm 0.25$ & 1.0(111) \\
       & 4.992  & 5.568 & $38^{+1.5}_{-1.6}$ & $11^{+0.5}_{-0.5}$ & 0.84(112) & 5.39 &$0.61\pm 0.28$ & 1.1(110) \\
       & 5.568  & 6.144 & $36^{+1.5}_{-1.5}$ & $11^{+0.5}_{-0.5}$ & 1.11(112) & 5.23 &$0.6\pm 0.33$ & 1.17(110) \\
       & 9.792  & 10.496& $39^{+2.0}_{-1.9}$ & $7.7^{+0.4}_{-0.4}$ & 1.24(113) & 3.56 &$1.32\pm 0.40$ & 1.63(111) \\
       & 10.496 & 11.2  & $37^{+2.0}_{-2.0}$ & $6.8^{+0.4}_{-0.4}$ & 1.21(114) & 3.2  &$0.54\pm 0.23$ & 1.03(112)  \\
       & 14.464 & 15.04 & $50^{+2.3}_{-2.2}$ & $14^{+0.7}_{-0.8}$ & 0.83(112) & 5.16 &$0.62\pm 0.24$ & 0.83(110)  \\
       & 15.04  & 15.616& $49^{+2.2}_{-2.2}$ & $13^{+0.6}_{-0.7}$ & 1.11(115) & 4.851& $0.62\pm 0.27$ & 1.4/109\\
       & 15.616 & 16.256& $46^{+2.2}_{-2.2}$ & $12^{+0.6}_{-0.7}$ & 0.9(112)  & 4.6 &  $0.61\pm 0.27$ & 1.4(109)\\ 
      &  &  &  &  &  &  \\ 
910927 & 0.000  & 0.704 & $63^{+3.8}_{-3.8}$ & $7^{+0.6}_{-0.5}$ & 0.82(107) & 2.26 & $0.62 \pm0.26 $ & 0.96(105) \\
       & 0.704  & 1.28  & $53^{+2.0}_{-2.4}$ & $12^{+0.6}_{-0.6}$ & 0.91(109) & 4.05& $1.02 \pm 0.28 $ & 0.93(107) \\
       & 1.28   & 1.792 & $46^{+1.7}_{-1.7}$ & $12^{+0.6}_{-0.6}$ & 1.3(107)  & 4.9 & $1.04 \pm 0.25 $ & 1.13(105) \\
       & 1.792  & 2.304 & $42^{+1.4}_{-1.3}$ & $12^{+0.5}_{-0.6}$ & 1.23(105) & 5.35& $1.24 \pm 0.24$ & 0.9(103)\\
       & 2.304  & 2.752 & $41^{+1.3}_{-1.4}$ & $14^{+0.5}_{-0.6}$ & 0.99(104) & 5.9 & $1.1 \pm 0.24$ & 0.8(102)\\
       & 2.752  & 3.2   & $37^{+1. }_{-1.2}$ & $15^{+0.6}_{-0.5}$ & 0.98(105) & 6.9 & $1.05 \pm 0.24 $ & 1.0(103) \\
       & 3.2    & 3.584 & $37^{+1.0}_{-1.2}$ & $17^{+0.6}_{-0.7}$ & 0.98(101) & 7.85& $1.12 \pm 0.25$ & 1.0(99)\\
       & 3.584  & 3.968 & $36^{+1.0}_{-1.0}$ & $17^{+0.6}_{-0.6}$ & 0.99(101) & 8.03& $0.98 \pm 0.19$ & 1.1(99)\\
       & 3.968  & 4.416 & $32^{+1.0}_{-1.0}$ & $13^{+0.5}_{-0.5}$ & 0.8(105)  & 6.9 & $0.83 \pm 0.2 $ & 0.8(103) \\
      &  &  &  &  &  & & & \\ 
970111 & 0.576  & 1.088 & $61^{+2.0}_{-2.0}$ & $11^{+0.8}_{-0.8}$ & 0.97(108) & 3.2  & $1.51\pm 0.4$ & 0.96(106)  \\
       & 1.088  & 1.6   & $56^{+2.3}_{-2.3}$ & $10^{+0.6}_{-0.6}$ & 1.15(105) & 3.23 & $1.31\pm 0.34$ & 1.07(103)  \\
       & 1.6    & 2.112 & $51^{+2.3}_{-2.3}$ & $9.3^{+0.6}_{-0.6}$ & 1.00(108) & 3.3 & $1.11\pm 0.28$ & 0.9(106)  \\
       & 2.112  & 2.58  & $49^{+2.4}_{-2.4}$ & $9^{+0.6}_{-0.6}$ & 1.43(105) & 3.32   & $1.02\pm 0.31$ & 1.2(103) \\
       & 2.58   & 3.008 & $48^{+2.4}_{-2.4}$ & $9^{+0.6}_{-0.6}$ & 0.98(104) & 3.3   & $1.6\pm 0.4$ & 0.9(102) \\
       & 3.008  & 3.456 & $49^{+2.4}_{-2.4}$ & $10^{+0.6}_{-0.6}$ & 0.93(104) & 3.6  & $1.02\pm 0.32$ & 0.8(102) \\
       & 3.456  & 3.904 & $46^{+2.1}_{-2.1}$ & $10^{+0.5}_{-0.5}$ & 0.97(102) & 4.0   & $1.85\pm 0.38$ & 0.7(100) \\
       & 3.904  & 4.288 & $46^{+2.0}_{-2.0}$ & $10^{+0.6}_{-0.6}$ & 0.95(104) & 4.06  & $1.83\pm 0.4$ & 0.9(102)\\
       & 4.288  & 4.672 & $40^{+1.9}_{-1.4}$ & $9.7^{+0.6}_{-0.6}$ & 1.16(100) & 4.27  & $1.25\pm 0.31$ & 1.2(98) \\
       & 4.672  & 5.056 & $42^{+2.0}_{-2.0}$ & $10^{+0.6}_{-0.6}$ & 1.13(98) & 4.3   & $1.34\pm 0.41$ & 0.92(96)\\
       & 5.056  & 5.44  & $43^{+2.0}_{-2.0}$ & $11^{+0.7}_{-0.7}$ & 0.96(104) & 4.7   & $0.79\pm 0.31$ & 1.07(102) \\
       & 5.44   & 5.76  & $44^{+2.0}_{-2.0}$ & $12^{+0.7}_{-0.7}$ & 0.85(101) & 5.0   & $0.84\pm 0.28$ & 0.82(99) \\
      &  &  &  &  &  &  \\  
\hline
\end{tabular} \end{center} }
\end{table*}
%_______________________________________________________________

As already mentioned the fits have been performed only on spectra with
$\alpha\gtrsim 0.5$. In order to determine the value of this threshold
in $\alpha$ we simulated spectra with a black body model assuming the
detector response of these bursts and a typical exposure time $\sim
0.2$ sec (i.e. the average integration time of the spectra presented
before). We then fitted these simulated thermal spectra with a
non--thermal model $N(E)\propto E^{\alpha} exp(-E/E_{0})$ where the
power law spectral index $\alpha$ was fixed at different values
between 0 (a flat photon spectrum) and +1 (which best approximates a
thermal spectrum). The residuals of these fits indicated that for
$\alpha<0.5$ the model spectrum (at low energies) deviates
systematically from the data (i.e. residuals different from 0) at more
than $2\sigma$. For this reason we selected $\alpha=0.5$ as the
threshold for fitting the spectra with a black body model.

The criteria applied for the goodness of the fit with the black
body model were:

\begin{itemize}
\item the reduced $\chi^{2}_{r}$ should be smaller than a fixed
reference value of 1.5.
\item the data--to--model ratio can be systematically and
significantly greater than 1 only at energies above the peak, as a
possible consequence of the presence of a supra--thermal or
multi-temperature spectral component (whose nature is not discussed
here).
\end{itemize}

The results of the fits for each spectrum are reported in
Tab. \ref{Black_body}: the start and stop time, with respect to the
trigger time, are given in col. 2 and 3. The model temperature $kT$
(keV) and its normalization (erg cm$^{-2}$ s$^{-1}$) are reported in
col. 4 and 5.  The value of the reduced $\chi^{2}_{r} / dof$ with
the corresponding degrees of freedom and the photon flux computed
from the fitted model in the 28--1800 keV energy range are given in
col. 6 and 7, respectively. Column 8 and 9 report the spectral index
($\alpha$) obtained by the fit with the non thermal (BAND or COMP)
models and the associated $\chi^{2}_{r}/dof$.

\begin{figure*}
\resizebox{17.5cm}{17cm}{\includegraphics[120pt,295pt][535pt,590pt]
{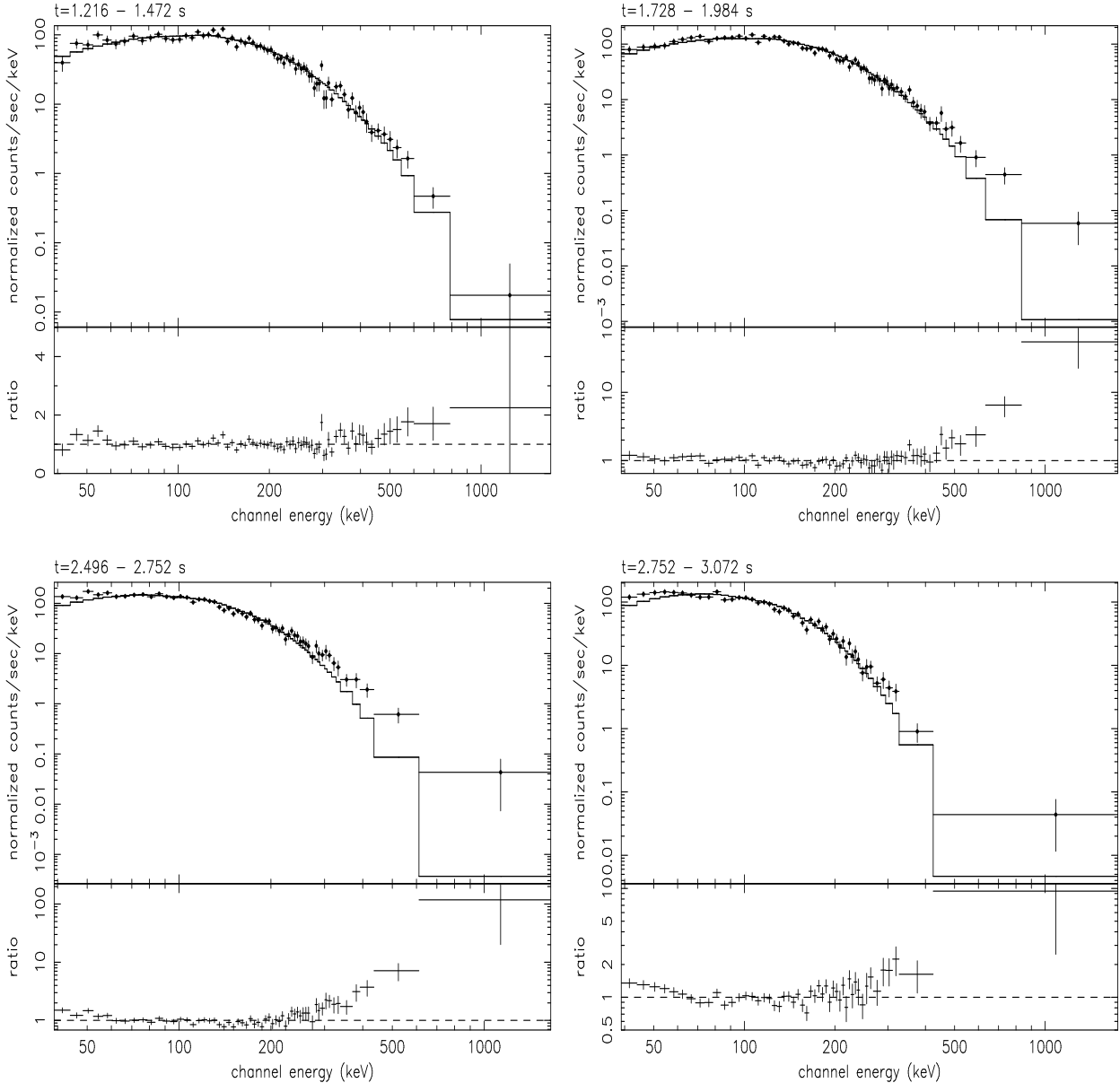}}
\caption[!]{ 
  Examples of black body fits for GRB 980306 (see
  Tab.\ref{Black_body}). The integration times are indicated for each
    plot. The data-to-model ratios are reported in the bottom panels.
  }
\label{fig2.1}
\end{figure*}

For illustration, we report in Fig. \ref{fig2.1} some examples of
fitted spectra for GRB 980306 and the corresponding data--to--model
ratios. The first two spectra are considered well fitted by the model
because $\chi^{2}_{r}=1.19$ and 0.9 for 111 dof, respectively, and the
ratio is systematically greater than 1 only at high energies. In the
other two cases the black body fits give less acceptable results
because the data--to--model ratio shows systematic deviations from 1
at low energies.

%___One column figure = Example spectrum _______________________
%____________________________________________________________________

In most of the bursts the best fits with the black body model are
obtained at the beginning: in the later stages (e.g. after the first
2.5 s in the case of GRB 980306) the low energy component softens 
and becomes incompatible with the black body model as the reduced
$\chi^2_{\rm r}$ and the data--to--model ratio indicate.

\begin{figure}
  \resizebox{\hsize}{!}{\includegraphics{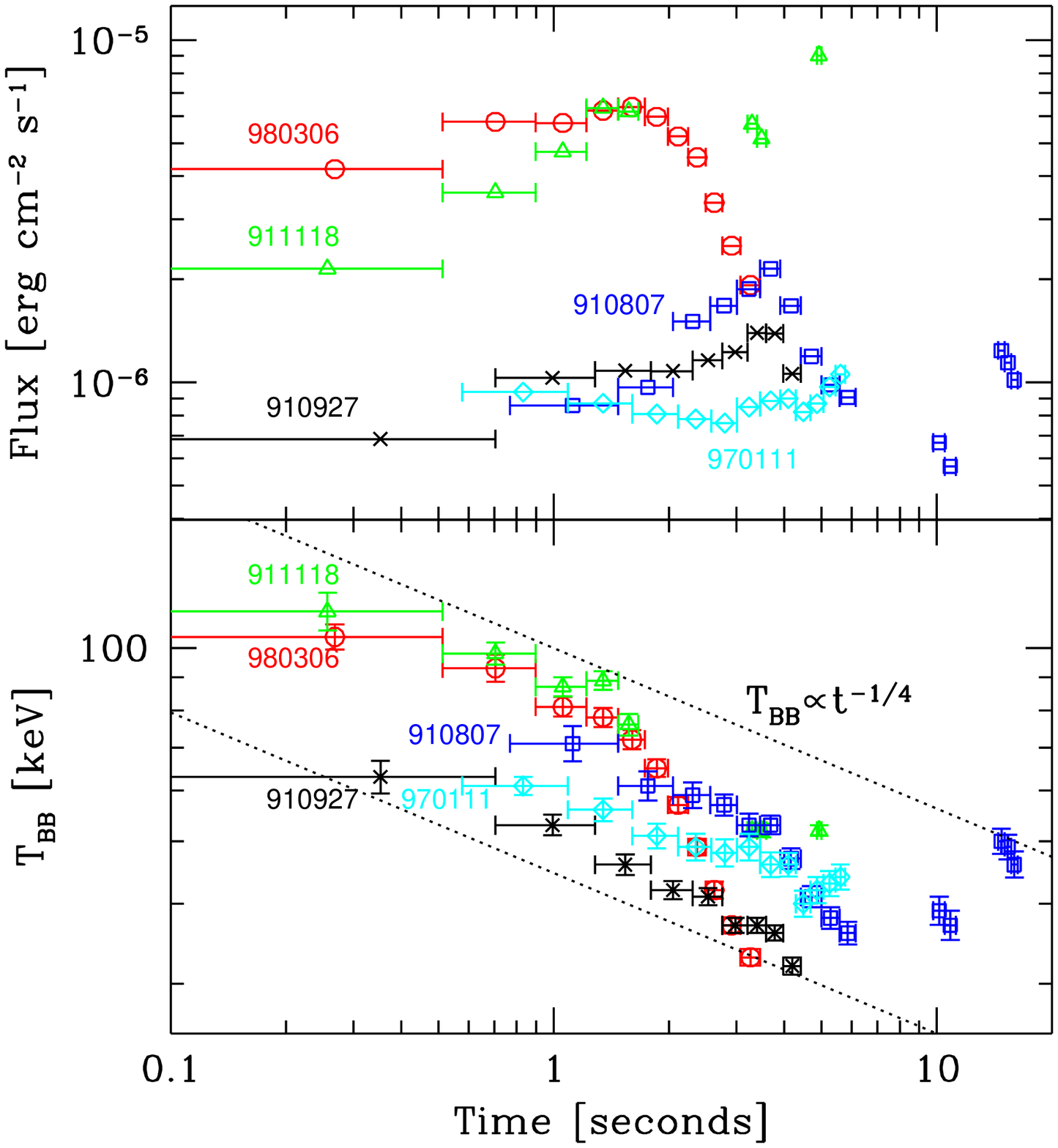}}
\caption[]{
  Total flux (top panel) and black body temperature (bottom panel) as
  a function of time for the spectra reported in Tab. \ref{Black_body}
  and considered acceptable in terms of residuals.  The dashed
  lines in the bottom panel correspond to $T_{BB}\propto t^{-1/4}$.  }
\label{fig2.2}
\end{figure}

The black body temperature and total flux (as derived by the model)
evolve in time as reported in Fig. \ref{fig2.2} for all the good black
body spectral fits. Note the decrease of the temperature with time in
almost all these bursts, which in the figure are compared with a
dependence $T_{BB}\propto t^{-1/4}$ (dotted line).  Although the
temperature decreases, the flux slightly increases or remains constant
in the first phases, and rapidly decreases thereafter.

Before concluding we note that although not publicly available, other
possible evidence of a black body emission from GRBs has been reported
in poster proceedings (Palmer D., private communication).  The
possible thermal black body spectrum of GRB970111 has also been
recently presented by Preece (\cite{Preece2001}): the spectrum
integrated over its first 5 s could be fitted by a black body with
$kT=55$ keV which is consistent with the average value of the black
body temperatures that we obtain from the fits of the time resolved
spectra covering the same time interval (Fig.\ref{fig2.2}).

%%___One column figure = 910807 evolution _______________________
%____________________________________________________________________

\section{Discussion}

GRB 980306 and 911118, together with GRB910807, 910927 and 970111
already reported in the literature, bring new constraints on the
extremely hard spectra that the emission process responsible for the
prompt phase must be able to satisfy.  These bursts represent the
cases with the hardest low energy power law spectral index and
significantly contribute in extending the $\alpha$ distribution
reported by Preece at al. (\cite{Preece2000}) and Ghirlanda et
al. (\cite{Ghirlanda}) to values greater than 0, up to $\alpha \sim
1.5$.  For the bursts studied here $\alpha$, i.e. the spectral index
of the low energy photon spectrum, remains positive for a major part
of the pulse/s, like in the case of GRB 911118 or GRB 970111, or for
its first rising part (GRB 980306 and GRB 910807).

In the following we consider several emission models proposed so far
to see if they can account for these hard spectra, except for the
photospheric and Compton drag model which will be discussed in 
Sec. 5.5.2 in relation to the quasi--thermal character of the initial
phases of these extremely hard bursts.

\subsection{Synchrotron emission}

{\it Optically thin synchrotron emission ---} The optically thin
synchrotron model, in its simplest formulation, can produce at most a
spectrum as hard as $\alpha=-2/3$ (Katz \cite{Katz}). Furthermore if
one considers the very short cooling timescales (much shorter than any
conceivable current exposure time) the predicted `cooled' spectrum has
$\alpha=-3/2$ (see also Ghisellini, Celotti \& Lazzati 2000).

\vskip 0.3 true  cm
\noindent
{\it Synchrotron self-absorption ---} One alternative, investigated
among others by Papathanassiou (\cite{Papathanassiou}), is that the
medium responsible for the synchrotron emission also absorbs these
photons. Also Granot Piran \& Sari (\cite{Granot}) have proposed
a model based on synchrotron self absorption by a stratified electron
population. In these cases it is possible to produce low energy
spectral slopes as hard as $\alpha=1$ (for a thermal or a power law
electron energy distribution with a low energy cut--off) or
$\alpha=1.5$ (for a power law extending to low energies).

However, for synchrotron self--absorption to be effective in the
typical BATSE energy range, a very high density of relativistic
electrons is necessary.  Assuming that the electron distribution
extends to low energies (this minimizes the required particle number)
with a power law distribution of index $p$ (i.e.  $N_{\rm
e}(\gamma)\propto \gamma^{-p}$, where $\gamma$ is the Lorentz factor
of the electrons), the synchrotron self--absorption frequency $\nu_t$
in the comoving frame, assuming $p=2$, is (e.g. Ghisellini \& Svensson
1991) $\, \nu_{\rm t} \, \simeq \, 2\times 10^{5} \nu_{\rm
B}\,({\tau_T / B})^{1/3} \, \, Hz,$ where $\tau_{\rm T}$ is the
Thomson optical depth and $\nu_{\rm B}=eB/(2\pi m_e c)$ is the Larmor
frequency.  In order to have $\nu_{\rm t} \sim 10^{16}$ Hz,
corresponding to $\sim \Gamma \nu_{\rm t}$ in the BATSE range, $\tau_T
B^2 \sim 5.7 \times 10^{12} \nu^3_{t,16}$. The Comptonization
parameter $y\sim \tau_{\rm T} \langle \gamma^2\rangle$ corresponding
to the required particle density, would largely exceed unity (making
inverse Compton process more efficient than synchrotron) unless
the magnetic field is extremely high ($B \sim 10^7$ G). This would in
turn imply a (magnetically dominated) fireball with isotropic
(equivalent) energy in excess of $10^{54}$ erg, using $E\sim 4\pi
R^{2} (B^{2}/8\pi) \Gamma^{2} c t_{burst}$ and assuming $\Gamma \sim
10^{2}$ and $R\sim 10^{13}$ cm. A magnetic field of $\sim
10^{7}$ G has been derived also by Crider \& Liang (\cite{Crider c})
from the fit of a simplified self absorbed synchrotron model to the
spectrum of GRB 970111.

\vskip 0.3 true  cm
\noindent
{\it Small pitch angles ---} Lloyd \& Petrosian (2000) have proposed
that in  rarefied, highly magnetized 
plasmas, turbulence can give origin to an anisotropic pitch
angle distribution of emitting particles.  In this case the 
optically thin synchrotron spectrum is modified at low energies
(compared to the standard $F(E)\propto E^{1/3}$), with a limiting
slope $F(E)\propto E$.  This can accommodate some, but not all, of the
hard spectra that we presented.  Note also that according to this
scenario to prevent strong cooling the particles have to be
re-accelerated, contrary to one of the basic assumptions of the
internal shock scenario, and in order not to be re--isotropized by
scattering, the inverse Compton process must be much less efficient
than the synchrotron one.

\subsection{Jitter radiation}

A variant of the standard synchrotron emission theory, the
jitter radiation (Medvedev \cite{Medvedev}), can justify a flat low
energy spectral slope. It originates as the emission of relativistic
electrons in a non uniform magnetic field with inhomogeneity
length scale smaller than the Larmor gyro--radius.  Medvedev
(\cite{Medvedev}) proposed a composite model for GRB spectra by
assuming the presence of a small scale magnetic field, which causes
the jitter radiation component, and a larger scale field producing a
standard synchrotron spectrum.
The composite spectrum has a broad bump in correspondence of the
jitter radiation characteristic energy, which depends only on the
magnetic field properties and not on the electron energy and if the
small scale field prevails, the spectrum has a limiting
$\alpha=0$.  This implies that only a minority of the hard spectra
reported here could be consistent with such a model.

\subsection{Compton attenuation}

Compton attenuation (Brainerd et al. \cite{Brainerd}) can produce a
low energy (20--100 keV) hard component because of the propagation and
scattering of the intrinsic spectrum by material (not participating to
the bulk flow) located between the fireball and the observer.  The
original high energy spectral shape can be unaltered because of the
decline of the Klein--Nishina cross section with energy, while at
lower energies, where the scattering process is more efficient, the
spectrum is hardened.  The most appealing features of this model are
that it could explain the clustering of the peak energy of bursts
around some hundred keV and produce a flat slope at low energies.
But in order for this to work, the Thomson optical depth of the
scattering material must be quite large (of the order of $\tau_{\rm
T}\sim10$) and this would smooth out the observed burst light curve by
smearing the variability on the smallest time scales. Furthermore, by
decreasing the transmitted (spectrally unaltered) flux, this mechanism
requires intrinsic powers greater than what we observe, exacerbating
the energy budget problem.

\subsection{Quasi--thermal Comptonization}

The above models are based on the assumption that the emitting
particles are highly relativistic, as a result of ``instantaneous''
acceleration at a shock front.  However, if the heating mechanism
operates on timescales of the order of the dynamical time (or the
light crossing time of a shell in the internal shock scenario), then
the heating and cooling processes could balance, leading to typical
energies of the electrons that are sub or only mildly relativistic
(Ghisellini \& Celotti 1999).  In this case the dominant radiation
process is Comptonization of seed soft photons by a quasi--thermal
particle distribution.

Different models assume different sources for the seed photons and
different parameters (i.e. typical size for the emitting region,
magnetic field, role of the photospheric radiation, and so on) within
the internal shock scenario.

Liang et al. (1997), among the first to propose such process for the
prompt emission, considered a relatively large emitting region
embedded in a relatively weak magnetic field.  This model was applied
to GRB 990123 by Liang (1999), who found quite extreme best fit values
for the required total power and number of emitting particles.
Ghisellini \& Celotti (1999) proposed that Comptonization acts on the
self-absorbed cyclo--synchrotron radiation produced by the same
quasi--thermal particles responsible for the multiple scattering
process, and pointed out the role that electron--positron pairs can
have in keeping the temperature (or the mean energy) of the particles
within a narrow range. On the same line M\'esz\'aros \& Rees (2000)
pointed out the importance of the residual photospheric emission as
source of soft photons.

The basic features of the Comptonization model is that, in the
quasi--saturated regime a Wien peak with its characteristic
$F(E)\propto E^3$ shape can form at high energies (i.e. for $h\nu \sim
kT$).  In principle this can thus explain very hard spectra.  One
possible difficulty of this model is the slope at lower energies,
whose saturated value should be $\alpha=-1$. A further possible
problem for quasi-thermal Componization is that the temperature of the
emitting particles (in the comoving frame) is expected to be of order
of $\sim 50$ keV, which would lead to an observed $E_{peak}$ of few
MeV (Ghisellini \& Celotti 1999). This is a somewhat high value, also
in the light of the results presented in this work, which show that
even spectra time resolved on scales $\le$ 1 sec present a peak energy
of $E_{peak}\le 500$ keV.

\subsection{The possible thermal character of the initial phase}

The first 1--5 s of the emission of the bursts presented in this work
were found to be consistent with a black body spectrum with typical
temperatures initially around $\sim$ 100 keV and decreasing to 30 -- 40
keV.  After this initial phase a softer (non--thermal or
multi--temperature) character of the spectrum becomes dominant.
Clearly this behavior can be a powerful diagnostic. In particular it
seems to favour two among the proposed models, namely photospheric and
Compton drag emission, which both predict -- as recalled in Sec.  5.5,
5.6 -- a thermal spectrum during the first phases.  Here we briefly
discuss which physical constraints can be quantitatively gathered from
the data.

\subsubsection{Photospheric emission}

When the fireball is becoming optically thin, during the acceleration
or coasting phase, its internal energy is emitted with a black body
spectrum whose observed temperature is blue-shifted by a factor
$\Gamma$ with respect to the comoving one. The possible presence of a
photospheric component in the spectrum of GRBs has been theoretically
investigated by M\'esz\'aros \& Rees (\cite{Meszaros}).  Recently,
Daigne \& Mochkovitch (2002) have reconsidered this possibility,
finding quite tight limits on the model of hot fireballs accelerated
through internal radiation pressure, imposed by the absence of an
initial emission phase (or precursor) with a black body shape.

In this scenario we have two options: we can relate the initial phase
to a single fireball, in the process of becoming optically thin, or --
following Daigne \& Mochkovitch (2002) -- to an ensemble of $\cal{N}$
shells, each becoming optically thin at approximately the same
distance.  In both cases at transparency:
\begin{equation}
\tau_{\rm T}\, =\, 
{ E_{\rm f} \sigma_{\rm T} \over 
2\pi  \theta^2 R_{\rm t}^2 m_{\rm p} c^2\Gamma \cal{N} } \, \sim \, 1\ \  , 
\end{equation}
where $R_{\rm t}$ is the transparency radius of the fireball,
collimated into two cones of semi--aperture angle $\theta$, $E_{\rm
f}$ is the total fireball energy (a fraction $\epsilon_\gamma$ of
which is radiated as photons), and $\sigma_{T}$ is the Thomson
cross section.  Assuming that the surface of the fireball is emitting,
at $R_{\rm t}$, as a black body with a comoving temperature
$T^\prime=T_{\rm obs}/\Gamma$, we have
\begin{equation}
\langle L_{\rm BB} \rangle \, =\, 
 {\epsilon_\gamma E_{\rm f}\over t_{\rm BB}} \, =\, 
2\pi\theta^2 R_{\rm t}^{2} \sigma \left({T_{\rm obs}\over \Gamma}\right)^4\,
\Gamma^{2} \ \ \ , 
\end{equation}
where $\langle L_{\rm BB} \rangle$ is the average black body
luminosity in the observer frame for the time $t_{\rm BB}$ and
$\sigma$ is the Stefan-Boltzmann constant.  Both eq. (2) and (3) show
a $E_{\rm f}\propto \theta^2R_{\rm t}^2$ dependence, allowing to solve
for the bulk Lorentz factor:
\begin{eqnarray}
\Gamma\ &=& \ 
\left( {\sigma_{\rm T} \sigma\,  \over m_{\rm p} c^2} \,\,
{T^4_{\rm obs}\, t_{\rm BB} \over \epsilon_\gamma {\cal N}} \right)^{1/3} \, 
\nonumber \\  
\ \        &\sim&\  3\times 10^3 T_{\rm obs, 9}^{4/3}\,  t_{\rm BB}^{1/3}\, 
\epsilon_\gamma^{-1/3} {\cal{N}}^{-1/3}
\end{eqnarray}
with $T_{\rm obs}= 10^9 T_{\rm obs, 9}$ K (corresponding to the 
observed black body peak of $\sim$ 100 keV).  We stress that this
estimate of $\Gamma$ is independent of the degree of collimation of
the fireball.

In the case of a single shell $t_{\rm BB}$ is the 
time needed for the fireball to become transparent, implying
\begin{equation}
R_{\rm t}\, =\, c t_{\rm BB} \Gamma^{2} \, \sim \, 3\times 10^{17} 
T_{\rm obs, 9}^{8/3}\,  t_{\rm BB}^{5/3}\,  \epsilon_\gamma^{-2/3}\,\, 
{\rm cm}.
\end{equation}
Therefore, if the emission comes from a single shell, the transparency
radius is very large, implying unreasonably large values for the
luminosity and energy, i.e. from eq.(3), 
$ E_{\rm f}\sim 3\times
10^{58}\, \theta^2_{-1} T_{\rm obs, 9}^{20/3}\, t_{\rm BB}^{11/3} \,
\epsilon_\gamma^{-5/3} \, \, \, {\rm erg}, $ 
where $\theta$ is
expressed in units of 0.1 radians.

If, instead, the observed emission
comes from a series of shells the transparency radius can be much
smaller, allowing a ``reasonable" luminosity of the photosphere: for
$E_{\rm f}=10^{51} E_{\rm f, 51}$ erg we have
\begin{equation}
R_{\rm t}\, \sim \, 5\times 10^{13} {E_{\rm f, 51}^{1/2} 
\over \theta_{-1} {\cal{N}}^{1/3}} \,
\left( {\epsilon_\gamma \over t_{\rm BB} T_{\rm obs,9}^4} \right)^{1/6}
\,\, {\rm cm}.
\end{equation}

We conclude that a series of shells (of total energy $E_{\rm f}$) each
becoming transparent at $R_{\rm t}$ can account for the observed black
body emission.  The case of $\cal{N}$ shells for large ${\cal N}$ is
clearly equivalent to a continuous or quasi--continuous flow.  

As Daigne \& Mochkovitch (2002) pointed out, and in agreement with our
findings, the photospheric radiation is likely to be visible only
during the first phases of the burst light curve, as long as the
optical depth of the material ahead of the shell which is releasing
its thermal radiation is negligible and before internal shocks take
over.

\subsubsection{Compton drag}

If the circum-burst environment is characterized by quite a large
photon density, as is the case of bursts following a supernova
explosion, or for fireballs produced in the matter--evacuated funnel
of an hypernova, there can be a strong interaction between these seed
photons and the fireball itself, as postulated in the so--called
Compton drag model.  The ambient photon energy is boosted by the
factor $\Gamma^2$ at the expense of the fireball kinetic energy
(Lazzati et al. \cite{Lazzati 2000}).  If the funnel or the young
supernova remnant are characterized by a single temperature $T_{\rm
SN}$, and if the fireball does not decelerate, the emitted spectrum is
a black body at a temperature $\sim 2 \Gamma^2 T_{\rm SN}$.  If
instead the seed photons have a range of temperatures (as likely to be
in the case of a funnel, hotter in the central parts), and/or the
process is so efficient to decelerate the fireball, than the final
spectrum will be a superposition of the locally produced black body
spectra, as calculated by Ghisellini et al. (\cite{Ghisellini}).  The
resulting spectrum can indeed resemble the very hard spectra found, at
least for the initial part of these bursts.  In fact, once the first
shells have swept up the seed photons, the efficiency of Compton drag
is greatly reduced, since the time to replenish the circum--burst
environment with seed photons is typically longer than the duration of
the burst itself.  On the other hand the Compton drag model favours
the formation of internal shocks, since after the first shells
deceleration, the subsequent ejecta can more easily collide with them
and produce shocks.  Then another radiation source can become
efficient after the first phase (lasting $\sim$1 to a few seconds).
This possibly also explains the hard to soft evolution: at first the
spectrum, due to Compton drag, is very hard at low energies
($F(E)\propto E^{2}$), whereas at later times the spectrum produced in
internal shocks through other processes becomes softer ($F(E)\propto
E^{1/3}$ or $\propto E^{0}$).

Let us then assume that the time interval for which the black body
lasts corresponds to the emission from a single shell and that
the emission peaks when the shell becomes transparent (eq. 2),
implying $R_{\rm t} \sim ct_{\rm t}\Gamma^2$.

The total energy $E_{\rm CD}$ due to the Compton drag process,
produced in the observed time $t_{\rm t}$, is of the order
\begin{eqnarray}
E_{\rm CD} \, & \sim &\, 2\pi\theta^2 R^3_{\rm t} (2 \Gamma^2) 
\, a \left({T_{\rm obs} \over 2\Gamma^2}\right)^4\,  \nonumber \\
&\sim& \,
 3 \times 10^{51}\theta^2_{-1} t_{\rm t}^3 T^4_{\rm obs,9}\, \,\, {\rm erg}
\end{eqnarray}
where $a=7.56 \times 10^{-15} {\rm erg\, cm^{-3} K^{-4}}$ is the
radiation constant.  In this form $E_{\rm CD}$ does not depend on
$\Gamma$ (although of course $T_{SN}$ does).
We can then derive the transparency radius (from eq.(2) and (8)) and
the bulk Lorentz factor:
\begin{eqnarray}
R_{\rm t}\, &=& \left({\sigma_{\rm T} c^{3/2} t_{\rm t}^{7/2} aT^4_{\rm obs}
\over 8 \epsilon_{\gamma} m_{\rm p}}\right)^{2/5} \nonumber \\
&\sim& \, 3\times 10^{14}\, T_{\rm obs, 9}^{8/5}\, t_{\rm t}^{7/5} \,
\epsilon_{\gamma}^{-2/5}
\,\, {\rm cm}
\end{eqnarray}
\begin{equation}
\Gamma\, =\, \left( R_{\rm t} \over c t_{\rm t}\right)^{1/2}
\, \sim \, 100 \,T_{\rm obs, 9}^{4 /5}\, t_{\rm t}^{1/5}\, \epsilon_{\gamma}^{-1/5}
\end{equation}
for a temperature of the seed photons corresponding to $T_{SN}
\sim 5\times 10^4$K.

These simple (and rough) estimates indicate that the Compton drag
model is viable. 

\subsubsection{Time evolution}

Although the limited number of time resolved spectra and bursts do not
allow to search for a general evolutionary behaviour, we can briefly
comment on the spectral and dynamical evolution of the initial thermal
phase of these 5 bursts.  As reported in Fig. \ref{fig2.2} in all the
5 bursts, the black body temperature decreases with time (bottom
panel) and its evolution is intriguingly close to $T_{\rm obs}\propto
t^{-1/4}$ for the first few seconds, and then drops. In the same time
interval the luminosity ($L_{BB}$) remains constant or mildly
increases (top panel of Fig. \ref{fig2.2}).  We therefore observe, at
least in the first phase, a decrease in the observed temperature
without a corresponding decrease in the observed black body flux.  We
plan to examine more deeply these observational behaviour in a future
work: here we just briefly comment on it in relation to the scenarios
discussed above.

In the context of the photospheric model discussed in Sec. 5.7.1, 
the black body evolution could be explained by successive shells 
having an increasing baryon loading and smaller Lorentz factor, 
thus becoming transparent at increasing distances $R_t$.
This can cause the observed temperature to decrease
(because the Lorentz factor is smaller) without a decrease
in flux (because the radius is larger).

In the context of the Compton drag model, a decrease in the observed
temperature can be due to the deceleration of the fireball and/or a
decreasing temperature of the seed photons with distance.  If we
consider the luminosity due to the Compton drag process (Eq. (8)), we
have $L_{\rm CD} \propto E_{\rm CD} /t \propto t^2 T_{\rm obs}^4$,
which predicts $L_{\rm CD} \propto t$ if the time behavior of the
observed temperature is indeed $T_{\rm obs}\propto t^{-1/4}$.  A
weaker dependence of the observed flux on time can occur if the
fireball is becoming transparent (i.e. only a fraction $\tau_{\rm T}$
of the seed photons can be scattered) with $\tau_{\rm T}$ decreasing
with distance and/or if the fireball distance becomes larger than the
typical dimension occupied by the seed photons.

\section{Conclusions}

We have presented the spectral evolution of GRB 911118 and of a new
case of hard burst, GRB 980306, together with some other hard bursts
already reported in the literature.  Their low energy spectral
component is harder than $N(E)\propto E^{0}$ for a considerable period
of their main peak emission and even after this initial phase their
spectrum remains harder than the synchrotron limit $E^{-2/3}$.  We
have applied different tests to verify the significance of the low
energy hardness, concluding that these results are indeed robust.

These GRB prompt spectra represent a challenge for the proposed
emission scenarios as shown through a comparison with the limiting
spectral shapes predicted by such models.  We pointed out the
difficulties that synchrotron emission, even including the effects of
self--absorption, small pitch angles particle distributions and
jittering, has in explaining spectra harder than $N(E) \propto E^0$.
Comptonization models also have difficulties, even if they are
consistent with very hard spectra in a limited range of energies.  Of
course, such a pruning of the forest of emission scenarios refers only
to the phases characterized by the harder emission, while it is
possible and likely that different processes dominate at different
stages of the prompt evolution and/or in different GRBs (see also
below).

The main result of this study is the possible thermal character of the
first emission phase of all the bursts we considered and the findings
on the evolution of such a thermal phase. The conventional scenario,
of a fireball accelerated by its own internal pressure, indeed {\it
predicts} such an initial thermal character when the fireball becomes
transparent. It was indeed its absence in previously considered bursts
that led Daigne \& Mochkovitch (2002) to favour a rather cold fireball
scenario, where at least part of the acceleration was due to magnetic
forces.  In our bursts, on the contrary, the luminosity in the thermal
phase is a significant fraction of the total, therefore in agreement
with the hot fireball scenario.  However, the fireball could be cold
initially, and be heated later (but before it becomes transparent).
The heating agent could be magnetic reconnection (see e.g. Drenkhahn
\& Spruit 2002) or internal shocks occurring at early phases.  The
latter case demands that the typical bulk Lorentz factors of the
shells are small enough for the shells to collide before the
transparency radius.

Alternatively, a cold fireball could work if the thermal spectrum we
see is produced by Compton drag, in which circum-burst radiation is
boosted to high energies by the fireball bulk motion.

At present, we are not able to discriminate between the photospheric
and the Compton drag scenario.  However, a key difference between the
two scenarios is the fact that the seeds photons for the Compton drag
process can be ``used'' only by the first shells, because the time
needed to refill the scattering zone with new seeds exceeds the
duration of the burst.  Therefore observing blackbody emission for a
long time, or during the rising phase of two time resolved peaks would
be difficult to explain in terms of the Compton drag process.  We
finally note that emission at times greater than a few seconds can
well be due to other processes, possibly linked to internal shocks
starting to dominate at later phases.

\begin{acknowledgements}
We thank the anonymous referee for a careful and thorough work and
her/his constructive criticisms.  This research has made use of data
obtained through the High Energy Astrophysics Science Archive Research
Center On line Service, provided by the NASA/Goddard Space Flight
Center.  Giancarlo Ghirlanda and AC acknowledge the Italian MIUR for
financial support.
\end{acknowledgements}

\end{document}